\documentclass[psamsfonts]{amsart}

\usepackage{natbib}
\numberwithin{equation}{section}

\usepackage{graphics}
\usepackage[dvipdf]{graphicx}
\usepackage{epsfig}
\usepackage{url}

\usepackage{natbib}

\usepackage{epsfig}

\usepackage{amssymb}

\usepackage{pifont}

\usepackage{supertabular}

\begin{document}

\title[CLIMATE WARMING AND STABILITY OF COLD HANGING GLACIERS]{Climate warming and stability of cold hanging glaciers:\\ Lessons from the gigantic 1895 Altels break-off}

\author{J. Faillettaz}
\address{J. Faillettaz\\VAW, ETH Zurich, Laboratory of Hydraulics, Hydrology and Glaciology\\Switzerland} 
\email{faillettaz@vaw.baug.ethz.ch}
\url{http://www.glaciology.ethz.ch}
\author{ D. Sornette}
\address{DD. Sornette\\epartment of Management, Technology and Economics, ETH Z\"urich}
\address{Department of Earth Sciences, ETH Z\"urich}
\address{Institute of Geophysics and Planetary Physics, UCLA}
\curraddr{Department of Management, Technology and Economics,\\ ETH Z\"urich\\Switzerland}
\email{ dsornette@ethz.ch}
\author{M. Funk}
\address{M. Funk\\VAW, ETH Zurich, Laboratory of Hydraulics, Hydrology and Glaciology\\Switzerland}
\email{funk@vaw.baug.ethz.ch}
%
%
\begin{abstract}

The Altels hanging glacier broke off on September 11, 1895. The ice volume of this
catastrophic rupture was estimated at $\rm 4.10^6$ cubic meters and is the largest ever observed ice fall event in the Alps. The causes of this collapse are however not entirely clear. Based on previous studies, we reanalyzed this break-off event, with the help of a new numerical model.
This model, initially developed by \citet{faillettaz&al2009} for
gravity-driven instabilities, was applied to this glacier. The model takes into account the progressive maturation
of a heterogeneous mass towards a gravity-driven instability, characterized 
by the competition between frictional sliding and tension cracking.
We use an array of sliding blocks on an inclined
(and curved) basal surface, which interact 
via elastic-brittle springs. A realistic state- and rate-dependent friction
law is used for the block-bed interaction.
We model the evolution of the inner material properties of the ice and its progressive damage eventually
leading to failure, by means of a stress corrosion law governing
the rupture of the springs.  The simulations indicate that 
a break-off event is only possible when the basal friction at
the bedrock is reduced in a restricted area,
possibly induced by the storage of infiltrated water within the glacier. This result
is in agreement with the fact that the collapse occurred after several hot summers.
Moreover, our simulations reveal a two-step behavior: 
(i) A first quiescent phase, without visible changes, with a duration depending on the
rate of basal changes; (ii) An active phase with a rapid increase of basal
motion over a few days. As a consequence, 
a crown crevasse is predicted to open within a few days (which was observed) prior to
the occurrence of the final collapse. The general lesson obtained 
from the comparison between the simulations and the available evidence
is that visible signs of the 
destabilization process of a hanging glacier, resulting from a progressive warming of
the ice/bed interface towards a temperate regime, will appear just a few 
days prior to the collapse.
\end{abstract}
\maketitle

\section{Introduction}

Icefalls pose a considerable risk to humans, settlements and infrastructures. 
The destructive power of this natural phenomenon is usually greater in winter
as it may drag snow and ice on its train. In the Alps, one of the most tragic
ice fall event occurred in 1965 in Switzerland, when a major part of 
Allalin glacier broke off and killed 88 employees of the Mattmark dam
construction site \citep{Rothlisberger1981,Raymond&al2003}.
Following this event, interest in the instability of hanging glacier grew
within the alpine glaciological community. In 1973, the first successfull
icefall prediction was performed by \citet{Flotron1977} and
\citet{Rothlisberger1981} at the Weisshorn glacier, which pose regulary a threat to the
village of Randa (Valais, Switzerland).  
Due to climatic variations, some hanging glaciers undergo rapid changes
leading either to isolated catastrophic events, or to new situations with no
historical precedent.
Direct measurements are difficult to perform on these steep, heavily crevassed
and avalanche-endangered glaciers. These measurements are often sparse and
fragmentary, and therefore difficult to interpret. They were always performed after clear signs of destabilization. The conditions
prevailing before an unstable state are thus unknown. The factors
responsible for the destabilization of large ice masses are the strength of the ice and the stresses in the zone of fracture. However, the
physics of the ice fracture and the feedback mechanisms between crevassing,
ice deformation and load distribution are complex and mostly unknown. Lack of
theory and sparse measurements make an accurate stability assessment difficult.

To cope with these difficulties, we developed a numerical model describing the progressive
maturation of a mass towards a gravity-driven instability,
which combines basal sliding and cracking (described in
\citet{faillettaz&al2009}). Our primary hypothesis
was that gravity-driven ruptures in natural heterogeneous material
are characterized by a common triggering mechanism
resulting from a competition between frictional sliding and
tension cracking. 

The present paper is devoted to the application of this general numerical
model to a particular gravity-driven instability: the breaking-off of hanging
glaciers. The gigantic
breaking-off of the Altels hanging glacier is an interesting case for several reasons:
(i) This break-off is the largest ever observed and recorded in the
Alps.
(ii)
This catastrophic phenomenon was very well documented as  \citet{Heim1895},
\citet{Forel1895} and \citet{DuPasquier1896} described, within the limit of their
knowledge, their observations related to the rupture of the glacier and made some photographs of the glacier
before and after the collapse. 
(iii) The causes of this glacier instability are not entirely understood, despite the
reanalysis made by \citet{Rothlisberger1981} of the data collected by \citet{Heim1895}, \citet{Forel1895},
\citet{DuPasquier1896}. However, hot summers prior to the event are suspected to
have triggered this break-off.
(iv) This is the only well documented break-off of a cold hanging glacier,
where a progressive warming of the ice-bed interface towards temperate conditions likely has driven the phenomenon.
(iv) Furthermore, the study of this break-off is of interest in the context of global climate warming. 
This study should help to better understand the processes leading to this
catastrophic phenomenon. Moreover, it should give new insights on the probable causes of this particular gravity-driven instability and reveal precursors of this event.

Section \ref{site} describes the study site and the qualitative description of
the rupture. Section \ref{numdes} describes how the model is implemented and
section \ref{numres} presents the main qualitative and quantitative results.

\section{Study site: Altels}
\label{site}
\subsection{Generality and description of the catastrophe}
\begin{figure}
\vspace*{2mm}
\begin{center}
\includegraphics[width=0.7\textwidth]{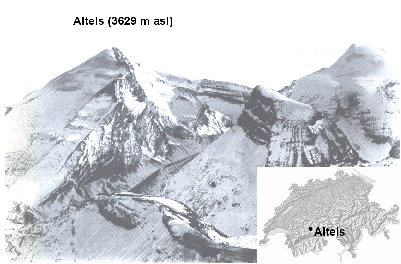}
\end{center}
\caption{\label{genview} Overview of the Altelsgletscher in November 25
  1894, ten months before
the glacier break-off. Photo P. Montandon from Englistliggrat, 2665 m asl}
\end{figure}

The Altels (Berner Oberland, Switzerland) is 3629 m above see-level and has a pyramidal shape. 
The north-western flank is 1500 m high and $35^\circ$ to $40^\circ$ steep
(Fig. \ref{genview}). It consists of relatively smooth malm limestone (Fig. \ref{bedrock}).
In the middle of the 19th century, this face was largely covered with an unbalanced ramp
glacier \citep{Pralong&Funk2006} located between 3629 m and 3000 m above sea level.

In the early morning of September 11, 1895, a large part of this glacier broke
off and tumbled down.  This ice fall lasted about one minute and the accompanying thunder
could be perceived in Kandersteg (about 10 km away). Many people thought it
was an earthquake. This catastrophic break-off was
carefully described and reported by \citet{Heim1895}, \citet{Forel1895},
\citet{DuPasquier1896} and later on by \citet{Rothlisberger1981}. 

The volume of the break-off was estimated at 4 Millions cubic meters,
which is the largest known glacial fall in the Alps. 
The resulting ice avalanche ravaged the
high mountain pasture situated underneath called the Spittelmatte and caused the
death of 6 persons and 170 cows. Four huts and large forest parts
were also destroyed and great quantities of cheese ruined (which was an
economic disasters for the bereaved families).
Due to its huge velocity \citep[i.e. 430 km/h][]{Heim1895,Rothlisberger1981},
this avalanche pilled 300 m up on the opposite slope towards the
\"Uschinengrat (Fig. \ref{scketch}). An area of about one square km of the pasture
was buried under a 3 to 5 meters thick ice layer.
As \citet{Forel1895} reported, a similar event had already occurred at the
same place in 1782, killing 4 persons and hundreds of domestic animals \citep{Raymond&al2003}. 
\citet{Forel1895} pointed out that, at that time, the summer was warmer than usual.
Nowadays, this won't happen any more because the Altels glacier has almost disappearred (a tongue remains on its left side, which will melt away in the near future, Fig. \ref{bedrock}). 

\begin{figure}
 \begin{tabular}{rl}
 \begin{minipage}{41mm}
\includegraphics[width=\textwidth]{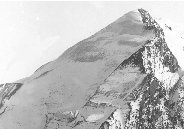}
 \end{minipage} &
 \begin{minipage}{41mm}
\includegraphics[width=\textwidth]{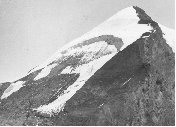}
 \end{minipage}
 \end{tabular}
\caption{\label{Altels1894} Altels glacier before and after its break-off (Photo
  P. Montandon, 25 November 1894 and 15 September 1895; Archiv des Alpinen Museums Bern)}
\end{figure}    

\begin{figure*}
\includegraphics[width=\textwidth]{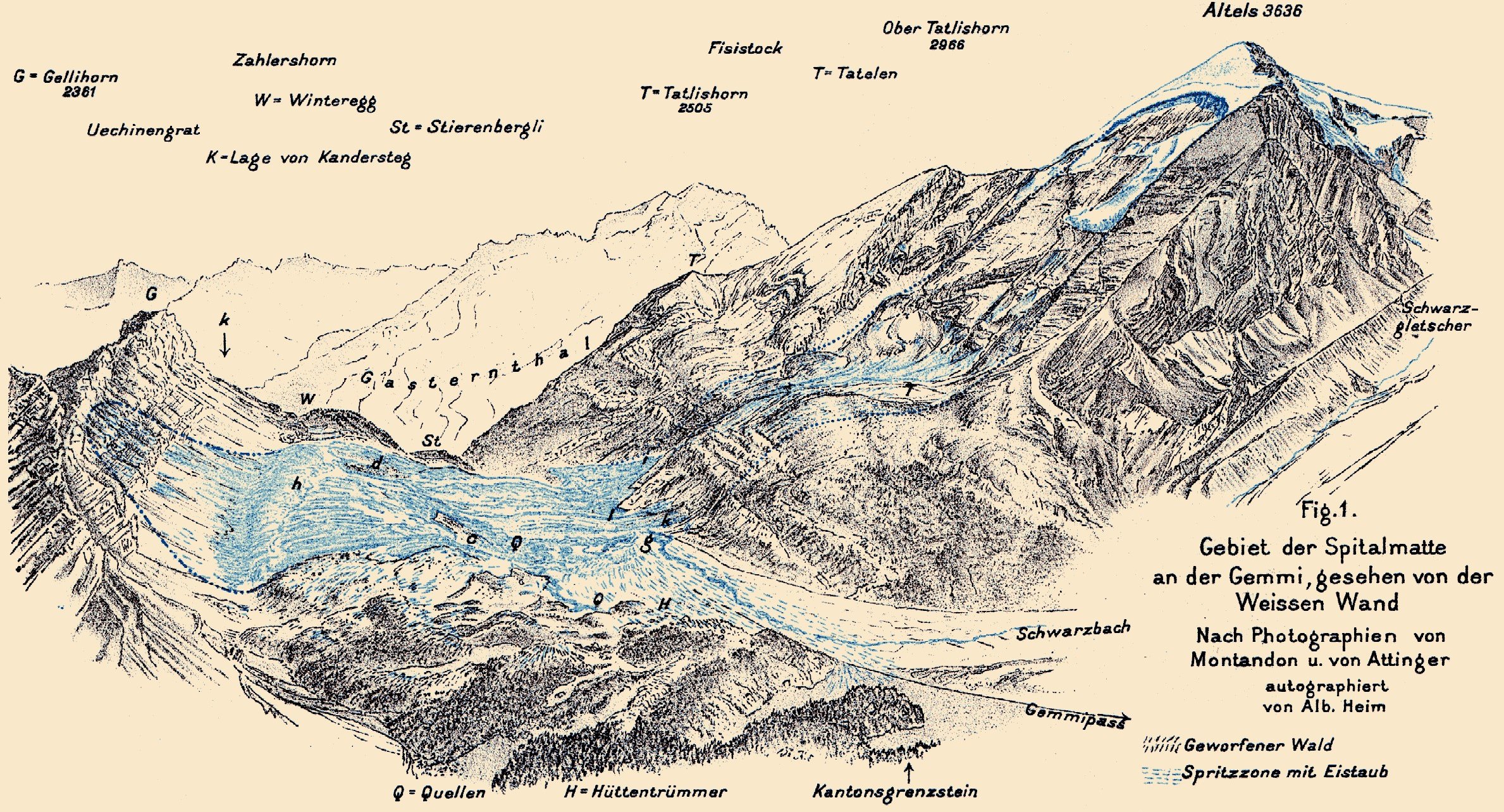}
\caption{\label{scketch} General overview of the catastrophe (after Heim 1895).}
\end{figure*}

\begin{figure}
 \begin{tabular}{rl}
 \begin{minipage}{41mm}
\includegraphics[width=\textwidth]{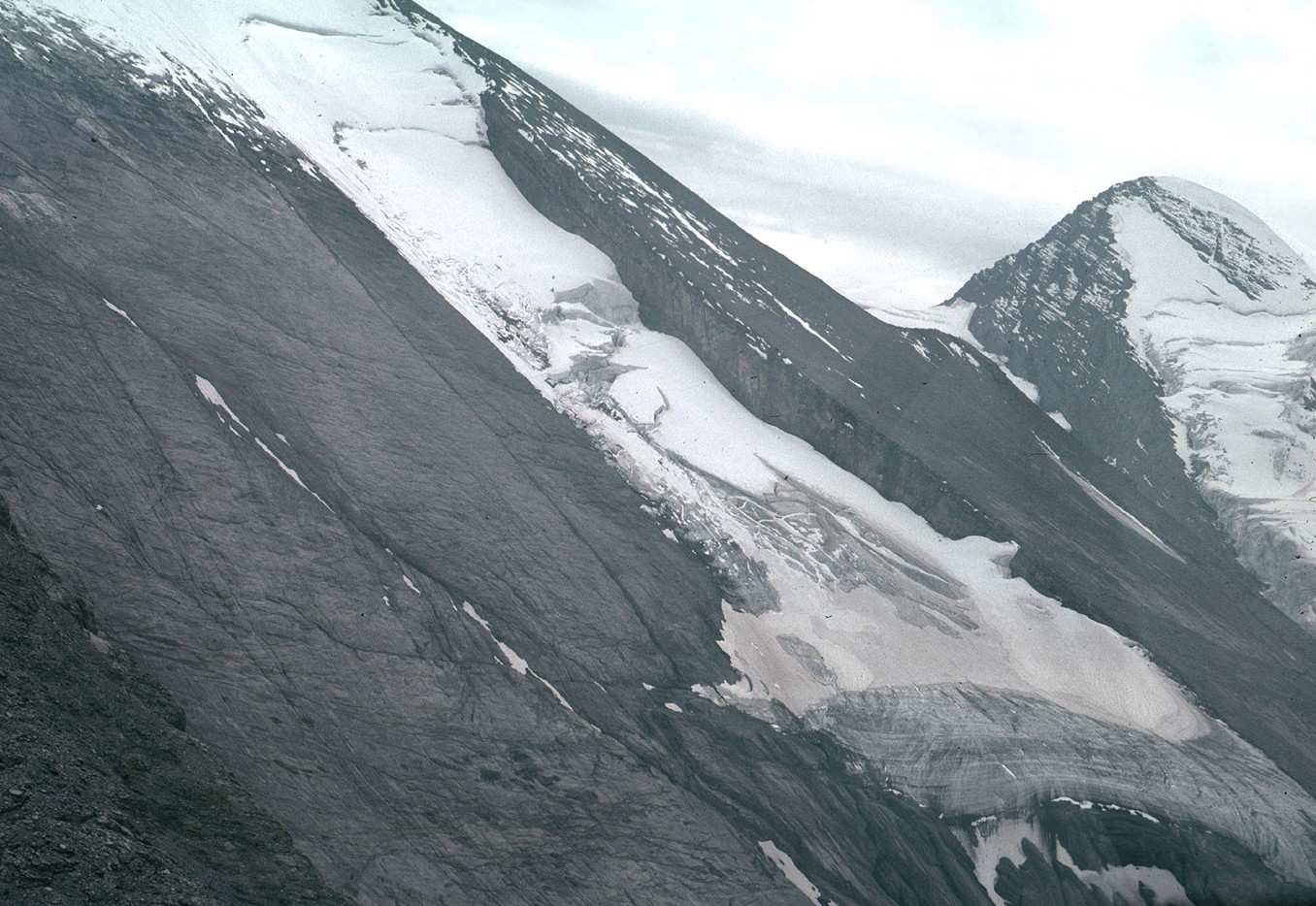}
 \end{minipage} 
&
 \begin{minipage}{41mm}
\includegraphics[width=\textwidth]{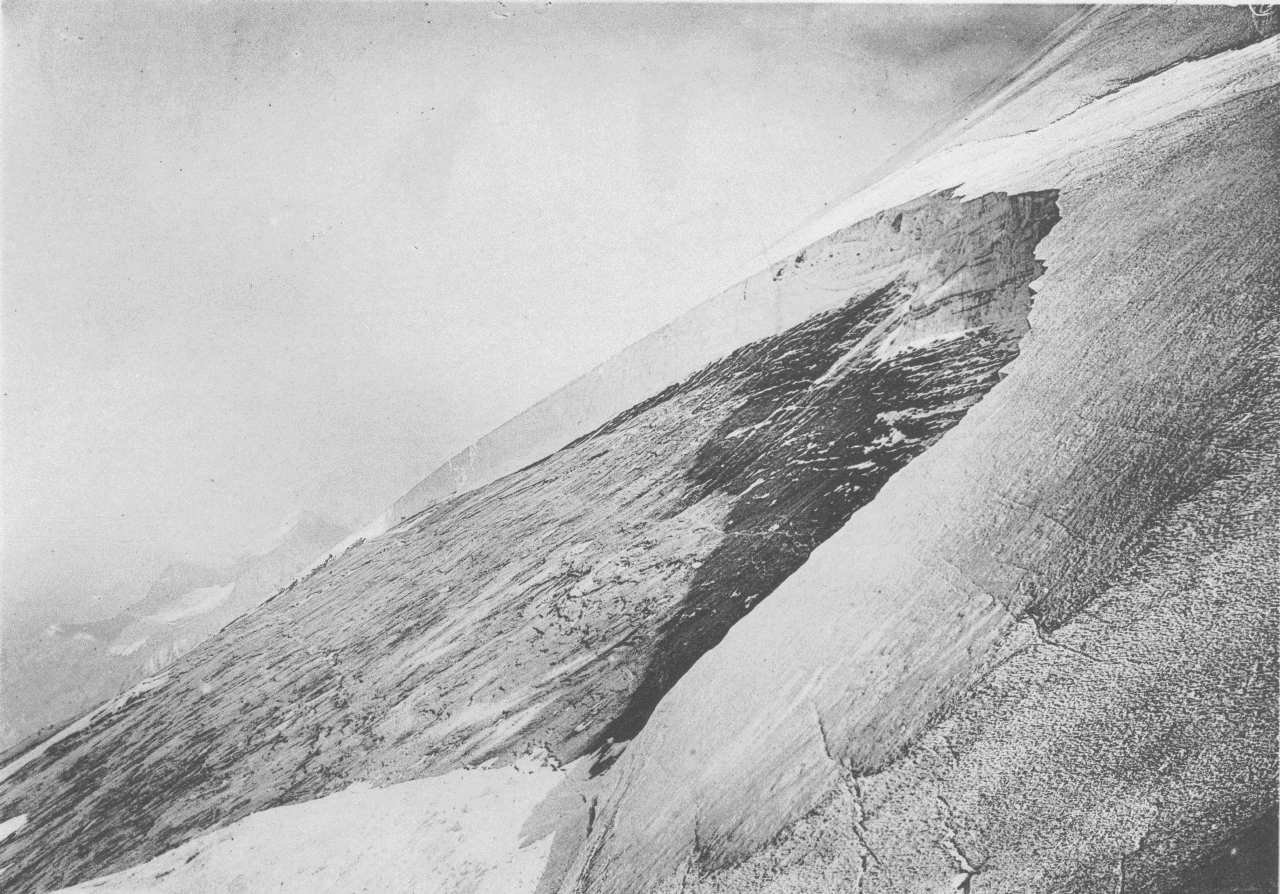}
 \end{minipage}

 \end{tabular}
 
 \caption{\label{bedrock} Link: Side glacier remaining in 1979. Bedrock consists of
  malm limestone. Photo from H. R\"othlisberger.
Right: Altels nowadays.} 

\end{figure}

\subsection{Rupture and probable causes}
\label{rupture}
The shape of the crown crack was a huge regular parabolic-like arch with a width of
about 580 meters (Figs. \ref{Altels1894} and \ref{bedrock}). 
The ice thickness at the crown crack was around 40 meters and 20
meters at the basis (Fig. \ref{bedrock}).
The finale rupture took place and propagated along the bedrock.

\begin{figure}
\begin{center}
\includegraphics[width=0.7\textwidth]{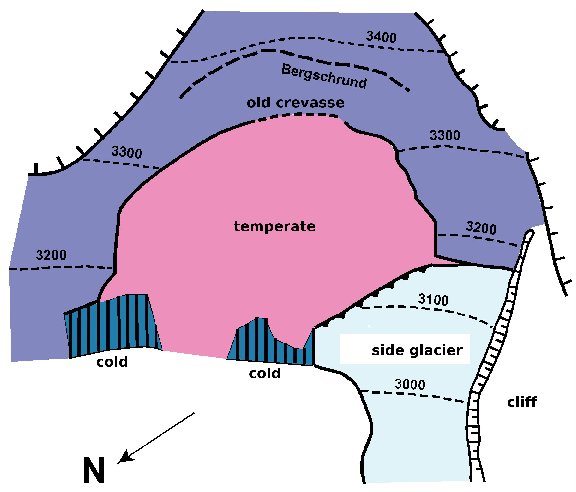}
\end{center}
\caption{\label{thermalcondition} Schema of the supposed thermal conditions at the
  bedrock and in the glacier before its break-off \citep[after][]{Rothlisberger1981}.}
\end{figure}

\citet{Forel1895} analysed the causes of the rupture mechanism. 
He pointed out the extremely hot previous summers as a possible link to
this catastrophe. This argument could explain why the glacier
was no longer mostly frozen within the bedrock and lost adhesion due to a thin
film of water between the bedrock and the ice. 
\citet{Rothlisberger1981} reanalyzed the documented field observations from
\citet{Heim1895,Forel1895,DuPasquier1896} to infer the
thermal conditions of Altels glacier before its break-off (Fig. \ref{thermalcondition}).
He argued, from Fig. \ref{Altels1894}  that:
\begin{itemize}
\item[(i) ]  Above the bergschrund, the glacier was frozen onto its bedrock and
  ice temperature should be below melting point.
\item[(ii) ] Below the bergschrund, ice and firn were temperate:
  it is possible to distinguish on Fig. \ref{Altels1894} a flat area covered by
  snow. By melting, this snow may have warmed up the ice and eventually melted water could have percolated through the glacier down to the bedrock.
\item[(iii) ] Further down, the snow cover lying on the
  glacier was very thin because of
  the strong wind erosion. \citet{Rothlisberger1981} argued that ice was likely be cold in this area, as the thermic isolation from the snow cover is less effective in colder season (Fig. \ref{Altels1894}). Moreover, he argued that melted water had no time to percolate and warm the glacier, as the slope is very large in this area.
\item[(iv) ] The glacier was frozen to the bedrock at the margins and temperate in the middle
  (Figs. \ref{Altels1894} and \ref{bedrock}). At the margins, some
  remaining ice can be recognized (white spots), indicating that the glacier
  was still frozen on the bedrock. 

\end{itemize}

\citet{Rothlisberger1981} analysed the retaining forces of the hanging glacier
and the possible causes of the break-off. 
\begin{itemize}
\item[(i) ]Adhesion to the bedrock: Given the dimensions of the
  unstable ice slab compared to its thickness, the basal properties should play a crucial
  role on the global stability. The existence of temperate conditions
  at the interface between bedrock and the glacier is a destabilizing process
  that favours sliding of the glacier on its bed. When the glacier slides
  over bumps or bedrock asperities, cavities forms above the bedrock leading to a lost of adhesion of the glacier on the bedrock. 
Moreover, melting and high water pressure at the interface
  favour and accelerate the enlargement of the cavities. This could lead to a 
  destabilization of the glacier in a very short time (from 1 day to 1
  week). Precursors of such destabilization process would be hardly detectable as such
  cavities can not be easily evidenced from the surface. 

\item[(ii) ]Southern support of the side glacier: The side glacier could also
  play an important role on the global stability of the hanging glacier. From
  Fig. \ref{genview}, it is possible to distinguish an opened crevasse on the side glacier
  prolongated into the Altels glacier ten months before the rupture. This
  could have formed because of sliding of the side glacier, which acts as a
  support for the whole glacier. Such a sliding phenomenon was already
  observed in 1895 and a sliding of 25/30 m/year was also observed in 1927/28 \citep{Rothlisberger1981}.

\item[(iii) ]Support of the frozen margins: It is possible that the lateral
  support of the frozen margins had a considerable role on the stability of the
  glacier, especially because of the important surface melting of the glacier two years
  before the rupture \citep[from photo
  in][]{DuPasquier1896}. \citet{DuPasquier1896} and \citet{Heim1895} analysed
  temperature data from different locations in Switzerland and found that the summer 1895 was exceptionnaly warm. 
  With the prevailing climate conditions, sufficient melted water could have been produced to weaken the basal frozen support.
  
\item[(iv) ]Traction at the crown crack (cohesion): As the shape of the crown crack is
  a nearly perfect arch, it seems that before the destabilization phase, the glacier
  was in a mechanical equilibrium. The crown crack was already slightly
  opened with a limited depth ten month before the rupture (see
  Fig. \ref{Altels1894}), indicating large tensile stresses.

\end{itemize}

\subsection{Temperature and precipitation before 1895}
As \citet{Heim1895,Forel1895} and \citet{Rothlisberger1981} pointed out that very warm summers (i.e. generating much more melted water than earlier) are suspected to have favoured the instability, we investigated the evolution of the climatic conditions at the Altelsgletscher before its 1895 break-off.
 A data base with homogenized continuous daily time series of temperature and precipitation since 1865 are available at the Federal Office of Meteorology and Climatology MeteoSwiss. The time series are corrected for systematic biases which may be due, for example, to the
relocation of the weather station or changing measuring techniques \citep{Begert&al2005}. 
A total of 12 MeteoSchweiz stations are available but we used the two closest stations, one located in Sion (35 km away) and other in Bern (60 km away).

Air temperature is known to be relatively well correlated over large distances \citep{Begert&al2005} and can therefore be extrapolated with confidence. We evaluated the temperature at 3000 m a.s.l. by applying a temperature gradient of -6$^\circ$C per 1000m on the mean daily values from the two meteo stations.
The extrapolation of solid precipitations are more problematic. As a first approximation, we computed the mean precipitations from the two stations and considered it to be solid when the extrapolated temperature at 3000 m was below 0$^\circ$C. 

Daily snow and ice melt rates can be assumed proportional to the positive degree day \citep{Hock2003}.
Because the albedo of ice is lower than for snow, the melting will be larger when the annual snow cover has disappeared. The rate of solid precipitation and the sum of the  Positive Degree Days (PDD) is an indication on the annual mass balance of the glacier, i.e. a small annual solid precipitation rates would lead to an earlier disappearance of the snow cover and therefore to higher melt rates because of a longer time with low albedo.

\begin{figure}
\begin{center}
\includegraphics[width=0.7\textwidth]{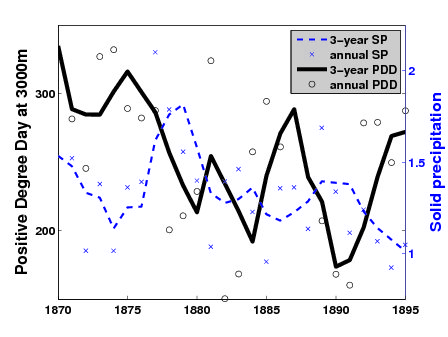}
\end{center}
\caption{\label{PDDWP} 3-year running mean of Positive Degree Days (solid black line) and solid precipitation (dashed line) between 1870 and 1895 at 3000 m a.s.l. at the location of the 1895 Altels break-off.
Annual values are also indicated (circles for PDD and crosses for solid precipitations).}

\end{figure}

Time series of Positive Degree Days (PDD) since 1870 (Fig. \ref{PDDWP}) indicated larger value in the 1870's and an increasing trend 5 years before 1895. Moreover, solid precipitation has decreased during this period indicating less winter accumulation. Five years before the break-off, the glacier experienced smaller annual mass balances (decrease of solid precipitation combined with a larger PDD). These results are compatible with the descriptions of \citet{Heim1895} and \citet{Rothlisberger1981}.

\section{Numerical modelling}
\label{numdes}

The aim of the present work is to reanalyse this event by applying a new numerical model designed for describing natural gravity-driven instabilities \citep{faillettaz&al2009}. This model allows us to test the different hypotheses previously published to explain the break-off of this glacier.

\subsection{Model description}
\label{shortdes}

We use a model describing the progressive maturation
of a mass towards a gravity-driven instability, which combines
basal sliding and cracking. Our hypothesis is that
gravity-driven ruptures in natural heterogeneous materials are characterized by a common
triggering mechanism resulting from a competition between frictional sliding
and tension cracking. Heterogeneity of material properties and dynamical
interaction seem to have a significant influence on the global behavior.

This numerical model is based on the discretization
of the natural medium in terms of blocks and springs forming a two-dimensional network
sliding on an inclined plane. Each block, which can slide, is connected to its
four neighbors by springs that can fail, depending on the history of
displacements and damage. We develop physically realistic models describing
the frictional sliding of the blocks on the supporting surface and the tensile
failure of the springs between blocks proxying for crack opening. Frictional
sliding is modeled with a state-and-velocity weakening friction law with
threshold. Crack formation is modeled with a time-dependent cumulative damage
law with thermal activation including stress corrosion. In order to reproduce cracking
and dynamical effects, all equations of motion (including inertia) for each
block are solved simultaneously.

The run-up to the
sliding instabilities can be described by applying a modern constitutive law of 
state-and-velocity dependent friction. This means that solid friction is not used as a
parameter but as a process evolving with the concentration of deformation and
properties of sliding interfaces. Cracking and fragmentation in the mass is accounted
for by using realistic laws for the progressive damage accumulation
via stress corrosion and other thermally activated processes aided by stress.

The present model improves the multi-block model of \citet{Andersen&al1997}
and \citet{Leung&Andersen1997} in two ways. First, we use a
state-and-velocity weakening friction law instead of 
a constant (or just state- or velocity-weakening) solid friction
coefficient. Second, rather than a static threshold for the spring failures, we model the 
progressive damage accumulation
via stress corrosion and other thermally activated processes aided by stress.
Both improvements make the numerical simulations significantly
longer but present the advantage of embodying rather well
the known empirical phenomenology of sliding and damage processes.
Adding the state and velocity-dependent friction law and time-dependent
damage processes allows us to model rather faithfully the 
interplay between sliding, cracking between blocks and the overall
self-organizing of the system of blocks. 

\begin{figure}
\begin{center}
\includegraphics[width=0.7\textwidth]{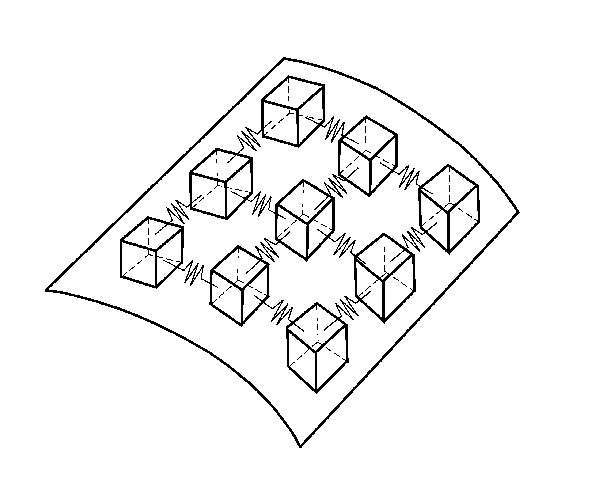}
\end{center}
\caption{\label{blocks} Illustration of the model constisting of spring-blocks resting on an inclined slope.
The blocks lie on an inclined curved surface and gravity is the driving force. Only
a small subset of the spring-block system is shown here.
}

\end{figure}

The geometry of the system of blocks interacting via springs and with 
a basal surface is depicted in Figure~\ref{blocks}. 
To sum up, the model includes the following characteristics:
\begin{enumerate}
\item Frictional sliding on the ground or between layers,
\item Heterogeneity of basal properties,
\item Possible tension rupture by accumulation of damage,
\item Dynamical interactions of damage or cracks along the sliding layer,
\item Geometry and boundary conditions, and
\item Interplay between frictional sliding and cracking.
\end{enumerate}

\begin{figure}
\label{algor}
\begin{center}
\includegraphics[width=0.7\textwidth]{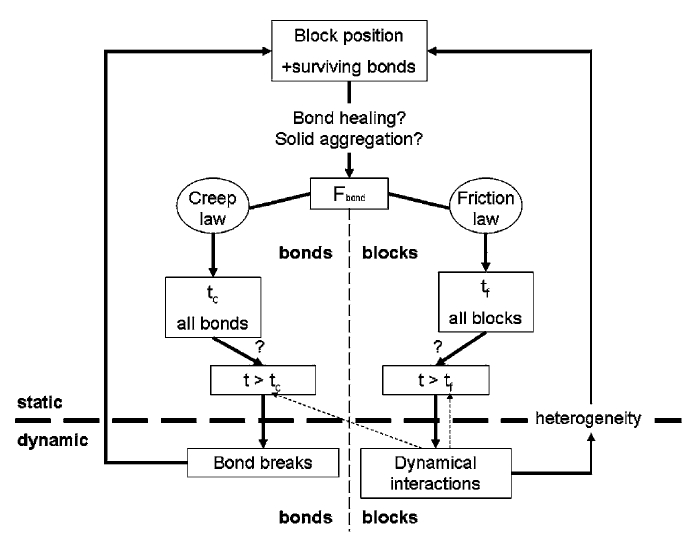}
\end{center}
\caption{\sf Schematic flowchart of this {\bf modified spring-block} model.
}
\end{figure}

The different steps describing how development of the instability is modelled are plotted
in Figure 8.
As explained previously, two phases have to be distinguished:
\begin{itemize} 
\item[(i) ]A quasi-static (quiescent) phase corresponding to the nucleation of block
  sliding and bond rupture.
\item[(ii) ]A dynamical (active) phase corresponding to the sliding phase of the blocks
  and of the failure of bonds.
\end{itemize}
To account for the changes of the surface characteristics after blocks have slid,  a
random state parameter $\theta_i$ taken between $\frac{\theta_0}{2}$ and
$\frac{3\theta_0}{2}$ is assigned to each stopping block.
In this way, the heterogeneity of the basal properties
can be reproduced and is sustained during the dynamic evolution of the system.

\subsection{Geometric parameters}
\label{geompar}
We first have to consider the geometric input parameters for modeling the
Altelsgletscher.
The glacier is discretized into a system of regular cubic blocks, whose weights 
remains constant during the numerical simulation.  The
application of this model to the time evolution of an unstable glacier 
implies that snow accumulation at the surface of the glacier is neglected, corresponding to a time
scale of one-two years. This assumption seems to be justified for
the Altelsgletscher as its northwestern face has a steep slope (from 35$^o$ to 40$^o$)
and is subjected to strong winds, drifting snow away. In addition, 
typical yearly additions of snow and ice are small compared with
the overall thickness (around 30 meters) of the glacier.
 
In order to obtain a realistic description of the damage and fragmentation
that may develop in the ice mass, we need a sufficiently large number
of blocks. As a compromise between reasonable sampling
and numerical speed, we use a model composed of $70 \times 70$ blocks for this
particular example.

It is possible to evaluate the size of the unstable part from the analysis of
Figure \ref{Altels1894} and from direct observations. 
Accordingly, the glacier surface area was approximately $4 \rm~km^2$,
with a mean ice depth of 30 m.  
As we consider a model composed of $70 \times 70$ blocks, each block 
corresponds to a discrete mesh with 30 m thickness, 30
m length and 30 m width. The weight of each block is approximately equal
to: $24.75 \times 10^6\;\; \rm kg$ (with a density of $917 \rm~kg.m^{-3}$).

\begin{figure}
\begin{center}
\includegraphics[width=0.7\textwidth]{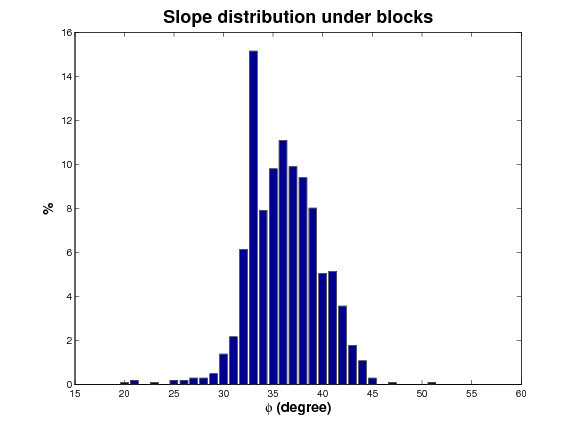}
\end{center}
\caption{\label{slope}Distribution of the slope of the bedrock at position of the blocks.}
 \end{figure}

\begin{figure}
\begin{center}
\includegraphics[width=0.7\textwidth]{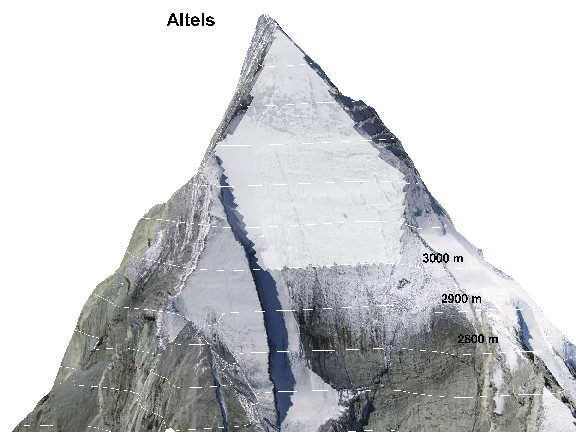}
\end{center}
\caption{\label{dem} Digital elevation model of the Altels.}
 \end{figure}

The slope $\phi$ of the bedrock ranges from $35^\circ$ (lower part) to $40^\circ$
(upper part) (see Fig \ref{slope}). To account for the curvature of the slope, we used a
digital elevation model (see Fig. \ref{dem}), supporting the blocks described above.

We now describe the two key processes in the model, the friction and
damage laws, that are applied to blocks and bonds respectively.

\subsection{Friction law between the discrete blocks and the basal surface}
\label{frictionlawchap}

As a first guess of the input friction parameters of this model, we choose
to start our investigations from the stablest case, i.e. cold
ice, stuck onto the bedrock. Obviously, this was not the case, as explained in section \ref{rupture}. Nevertheless, it is
a starting point for the  parametrization of the initial friction properties.

We first make the assumption that the friction coefficient
between the blocks and the underlying supporting sliding surface has constant
properties. This is a simplification as 
the water pressure at the glacier bed typically oscillates with a daily periodicity
\citep{Iken1981,Bahr&Rundle1996}. 
This assumption should however be valid in the case of a hanging glacier, for which,
without change of climatic conditions, the bedrock remains frozen and the ice
remains cold during the whole year.

In the model, the friction between the discrete blocks and the basal properties is described
with the following equation:
\begin{equation}
t_f\;=\;\frac{\theta_0}{\exp(\frac{\mu-\mu_0}{A})-1}~,
\label{frictionlaw}
\end{equation}
where $t_f$ is the time when the block starts sliding, $\mu_0$ is a constant
friction coefficient, A is a constant parameter depending on material
property and $\theta_0$ is the state parameter at steady state. $\mu$ is
evaluated on each block with the definition of the solid friction $\mu =
\frac{T}{N}$ where T are the tangential forces given by the position of its
connected neighbors and N is the normal component of the weight of the block.
Three parameters have to be evaluated to model the frictional processes within
the glacier: $\mu_0$, A and $\theta_0$. 
In the stablest case, the rupture should nucleate and propagate within the ice. 
More generally, the glacier would not slide on its bedrock, but damage is
concentrated in a sheared zone within the ice close to its base, resulting in
an observable displacement at the glacier surface.  Therefore, 
the friction law should represent the localization of shear within the ice, 
and the friction parameters should be taken to represent the physics
of friction between ice surfaces. 

Despite its importance, the topic of ice-ice friction has not yet been
extensively studied, especially at low velocity. It is known that
the ice-ice friction coefficient $\mu^{ice-ice}$ generally decreases with
increasing sliding velocities and ambient temperature. It is also known that
$\mu^{ice-ice}$ is relatively insensitive to both pressure and grain-size.
Generally speaking, ice is regarded as a material which exhibits very low friction in
sliding. However, at low sliding speed, the friction coefficient of ice can be
considerably higher \citep{Kennedy&al2000}.
These two different behaviors are generally explained by two physical mechanisms
depending on the sliding velocity regimes:
\begin{itemize}
\item[(i)] The first mechanism is the water lubrification mechanism (produced by frictional heat at the
sliding surface) working at sliding velocity above roughly 0.01 m/s. The water lubrification mechanism is characterized by the low viscous
resistance of water film produced by frictional heat at the sliding interface
\citep{Maeno&Arakawa2004,Barnes&al1971,Kennedy&al2000,Montagnat&Schulson2003}. 
\item[(ii)] The second mechanism is the adhesion and plastic deformation of
  ice at the friction interface, which is present at velocities lower than
  roughly 0.01 m/s \citep{Kennedy&al2000,Montagnat&Schulson2003,Maeno&Arakawa2004}.
 \end{itemize}
 \citet{Kennedy&al2000} found the ice-ice friction coefficient to vary by about
 an order of magnitude (0.8 to 0.1) over a wide range of velocity (from $10^{-6}$ to $10^{-1}$
m/s). At low velocities ($< 10^{-5}$m/s), no systematic effect of temperature on friction
was seen and tests performed at $-10^\circ \rm C$ at velocities ranging from $10^{-6}$ to
$10^{-1}$ m/s confirm that the nominal contact pressure has negligible effect
on the friction coefficient \citep{Kennedy&al2000}. 
In a recent study, \citet{Maeno&Arakawa2004} showed that the ice-ice friction
coefficient becomes as high as 1.0 at $10^{-6} \rm m/s$ for a temperature of
ice of $-10^o\rm C$, which is higher than values obtained by \citet{Kennedy&al2000}. 
They explained this large value by adhesion processes combined with ice
sintering at the sliding interface.

As far as we are concerned, high friction should be considered for the Altels
glacier for two main reasons:
\begin{itemize}
\item[(i)] Its characteristic velocity is indeed very low, around 1 to 10
cm/day, thus far below  0.01 m/s. 
\item[(ii)] The friction coefficient is independent of pressure. Thus, ice
  depth does not influence friction.

\end{itemize}
We thus choose a high value of the friction coefficient, namely the limit
value $\mu_o^{ice-ice} \approx 1$ given by \citet{Maeno&Arakawa2004}.

Once the block slides, the dynamics
is controlled by a kinetic friction coefficient, which is in general
smaller than the static coefficient $\mu_o^{ice-ice}$.
For low temperatures, \citet{Kennedy&al2000} found a relatively
constant value of the friction coefficient around $0.6$ for sliding
velocities below $10^{-4}$m/s, and rapidly decreasing values 
at velocities above $10^{-4}$ m/s to $0.1$. 
Because velocities above $10^{-4}$m/s are not expected in our model
describing the nucleation phase of the catastrophic rupture, we thus 
assume that the kinetic friction coefficient is $\mu_d=0.6$.  

Next, we turn to the coefficient $A$ (and $B$ since we have
made the simplifying assumption that $B=A$)  in the
rate- and state-dependent friction law given by equation (\ref{vxcxxzx}) \citet{faillettaz&al2009}: 
\begin{equation}
\mu(\dot{\delta}, \theta) = \mu_0 + A \ln {\dot{\delta} \over \dot{\delta}_0} + B \ln 
{\theta \over \theta_0}~,
\label{vxcxxzx}
\end{equation}
Here, $\dot{\delta}$ is the sliding velocity.

Laboratory experiments suggest that $A$ is smaller
than $\mu_0$ by typically one 
and sometimes up to two orders of magnitude \citep{Ohmura&Kawamura2007,
Scholz2002,Scholz1998}. As we do not have access to 
strong experimental constraints, we choose $A \approx 0.1$,
corresponding to one-tenth of the static friction coefficient.

The last parameter needed to parametrize the friction law is $\theta_0$ defined by \citet{faillettaz&al2009}
\begin{equation}
\label{theta0}
\theta_0\;=\;{D_c \over \dot{\delta}_0}.
\end{equation}
Here, $\dot{\delta_0}$ is generally interpreted as the initial low velocity
of a sliding mass, before it starts to accelerate towards its
dynamical instability. In the case of a glacier whose sliding velocity is 
typical small, i.e., of the order of centimeters per day, this suggests
that a rough correct estimation is $\dot{\delta_0} \approx 1 \;\rm cm.d^{-1}$.
$D_c$ can be interpreted as a characteristic slip distance over which
different asperities come in contact. It is
difficult to evaluate this value. The recent seismological literature
reports $D_c$ to lie in the range of tens of centimeters to meters for earthquakes
\citep{Mikumo&al2003,Zhang&al2003}. We arbitrary choose
$D_c \simeq 1 \;\rm m$.
Finally, inserting this value in equation (\ref{theta0}), we obtain $\theta_0=100 \;\rm days$. 

As explained in section \ref{frictionlawchap}, to account for the heterogeneity and
roughness of the sliding surface, the state variable $\theta_i$ is reset to a
new random value after the dynamical sliding stops. This random value should not be
choosen too low in order to prevent a block which just stopped sliding from switching
immediately in a new dynamical phase (i.e. $t_f = 0$). Thus, we assign 
$\theta_i=\nu . \theta_0$ with $\nu$ uniformly distributed between 0.5 and 1.5.

\subsection{Creep law}
\label{creeplaw}
As explained in section \ref{shortdes}, bonds are modeled as
linear springs in parallel with an Eyring dashpot. The springs
transmit the forces associated with the relative displacements
of the blocks. The spring stiffness has also to be evaluated in order
to reflect the elastic property of the bulk ice mass.
In continuous elasticity, Hooke's law of elasticity relates stress $\sigma$ and
strain $\epsilon$ via Young's modulus $E$: $\sigma=E \epsilon$.
This leads to the expression: $\sigma_{bulk}=E_{ice} \frac{\delta L}{L}$.
This stress is applied to a surface $S=L \times H$ (where $H$ correspond to the height and L the length of the surface), leading to an equivalent
force in the bulk equal to $F_{bulk}=E_{ice} \frac{\delta L}{L}\;S$.
A linear spring is subjected to forces given by $F_{bond}=K_{bond}\;\delta L$.
In order to find an equivalent behavior, these two forces have to be of the
same order, leading to a spring stiffness given by $K_{bond}=E_{ice} H$.
Usually, values for $E_{ice}$ are reported to be 9 GPa \citep{Petrovic2003,Petrenko&Whitworth1999}. 
However, there is a disagreement of an order of magnitude
between measurements of $E$ in laboratory (9 GPa) and from field observation
($\approx 1 \rm GPa$) as argued by \citet{Vaughan1995}.

Depending on the applied stress, ice has either a linear viscous behavior or a
non-linear viscous behavior.  
In glaciers, ice creep is usually described by a non-linear viscous deformation
called the Glen's flow law (see \citet{Hutter1983} and references therein). This law relates,
in steady-state conditions, strain rate and stress in the secondary creep regime. 
It is thus not possible with this law to describe the tertiary creep and the
time rupture of a bond.  Following
\citet{Nechad&al2005}, non-linear viscous behavior is introduced in our model by
the presence of the Eyring dashpot in parallel with a linear
spring of stiffness E. Its deformation e is governed by the
Eyring dashpot dynamics
\begin{equation}
\label{eyring}
\frac{de}{dt}=K \sinh(\beta s_{\rm dashpot})
\end{equation}
where the stress in the dashpot is given by 
\begin{equation}
s_{\rm dashpot}={s \over 1-P(e)} -Ee~,
\label{kthnmb}
\end{equation}
Here, $s$ is the total stress applied to the bond and $P(e)$ is the 
damage accumulated within the bond during its history
leading to a cumulative deformation $e$.
$P(e)$ can be equivalently interpreted as the fraction of representative
elements within the bond which have broken, so that the applied stress $s$
is supported by the fraction $1-P(e)$ of unbroken elements.
Following Nechad et al. [2005], we postulate the following dependence
of the damage $P(e)$ on the deformation $e$:
\begin{equation}
P(e)=1- \Bigg( \frac{e_{01}}{e+e_{02 }}\Bigg) ^\xi~,
\label{mgm,tbl;}
\end{equation}
where $e_{01}, e_{02 }$ and $\xi$ are 
constitutive properties of the bond material.

Finally, by combining the previous equations, \citet{faillettaz&al2009} ended up with a creep model that computes the critical time (i.e. failure of the bond) as a function of the stress experienced by the bond {\em s} given by:
\begin{equation}
\label{tcs}
t_c =\left\{ 
\begin{array}{ll}
{1 \over K} \;\exp(-\gamma \;s) & \textrm{if } s > s^\star\\
\to \infty & \textrm{if }  s \leq s^\star
\end{array} \right.
\end{equation}
where
\begin{equation}
\gamma=\frac{\beta e_{02}^\xi}{e_{01}^\xi}~.
\label{hy3ghte}
\end{equation}
and
\begin{equation}
s^\star=E\;\;\Big(\frac{e_{01}}{\mu}\Big)^\mu\;\;\Big(\frac{\mu-1}{e_{02}}\Big)^{\mu-1}.
\label{sstar}
\end{equation}
 Creep properties are defined by the parameters
$K$, $\beta$, $e_{01}$, $e_{02}$ and $\xi$, that we need to fix for
our simulations. 

We need to find the most appropriate
parameters to describe creep behavior of ice.
Natural glacier ice has a complex polycrystalline structure composed of
crystals of different sizes. But ice is a
fairly homogeneous material compared to fiber matrix composite. The
more homogenous a material,  the greater $\xi$. 
In the following, we set this value $\xi=10$ (which
means that ice is not very heterogeneous).

As Equation \ref{mgm,tbl;} shows, a fraction  $1-(e_{01}/e_{02})^\xi $ of all
present representative
elements undergo abrupt failure immediately after the stress is
applied. Because ice does not show significant damage
immediately after being loaded, this suggests to choose
$e_{01}=e_{02}$, so that this fraction is vanishing.

The other parameters describing the deformation of the Eyring dashpot
under an applied stress are $\beta$ for the non-linear term and $K$ for the
linear one. For this study, we choose 
$\beta\;=\;10^{-7}\; \rm Pa^{-1}$ and $K\;=\;10^{-3}\; \rm s^{-1}$.

\section{Numerical results}
\label{numres}

The aim of the numerical simulations is to test the different causes of the
rupture, summarized in \citet{Rothlisberger1981} and described in Section
\ref{rupture}.
In particular, we intend to provide answers to the following questions
\begin{itemize}
\item[. ] What is the influence of the glacier geometry on the dynamic of the instability?
\item[. ] Can such a rupture happen without changes of the basal properties?
\item[. ] How rapidly does  the instability develop?
\item[. ] Can we expect some precursors?
\end{itemize}

\subsection{Description of the simulation}

Blocks are distributed in a regular mesh along the slope, so
that bonds are initially stress free. At each time step, we evaluate, on each block, the slope from the digital
elevation model of the Altels area (see Fig. \ref{dem}).
This aims at mimicking the real topography of Altels glacier.
To determine the causes of the glacier collapse, we will test the different
contributions described by \citet{Rothlisberger1981} (section \ref{rupture}).

The table \ref{creeplawtab} summarizes the 
parameters used in our simulations.

 \begin{table}
\label{creeplawtab}
\begin{center}
\begin{tabular}{|cccccc|}
\hline
\multicolumn{6}{|c|}{{\bf Geometric parameters}}\\
n & $m_{block}$&$\phi$    \\
- & kg        & $^\circ$  \\
70 & $24.75 \times 10^6$&30 to 45\\
\hline
\multicolumn{6}{|c|}{{\bf Friction parameters}}\\
A  & $\theta_0$ & $\mu_0$ & $\dot{\delta_0}$\\
 - &    d       & -       &$cm.d^{-1}$\\
 0.1 & 100 & 1 & $10^{-3}$  \\
\hline
\multicolumn{6}{|c|}{{\bf Creep parameters}}\\ 
$E$&$\beta$ & $C \sim 1/K$  & $\xi$ &$e_{01}$&$e_{02}$\\
Pa&$\rm Pa^{-1}$ & $s$& -     & -   &   -    \\
 $10^9$   &$10^{-7}$ & $10^3$ & 10 & 0.003&0.003\\
\hline
\end{tabular}
\end{center}
\caption{Parameters used for the simulation. $n$ is the linear dimension of the lattice of blocks, which 
has a total of $n \times n$ blocks. The second row contains the list of parameters. The third row 
specifies the units of the parameters and in the fourth row the numerical values used
in the simulation are given.}
\end{table}
   

\subsection{Qualitative results}
\label{qualitativeres}

\subsubsection{Is the 1895 Altels collapse solely due to glacier thickening?} 
\begin{figure*}
 \centering
 \begin{tabular}{cc}

\begin{minipage}{0.48\textwidth}
\includegraphics[width=\textwidth]{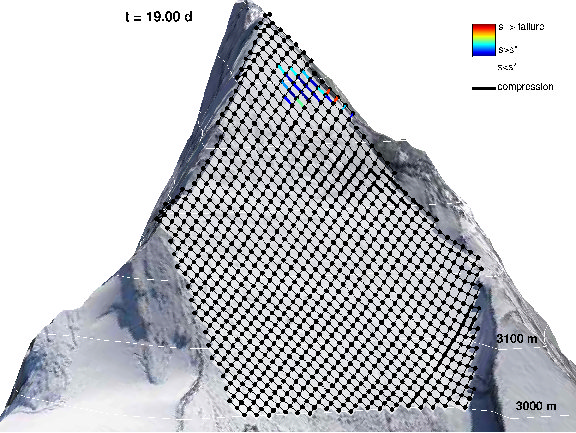}
 \end{minipage}
 &
 \begin{minipage}{0.48\textwidth}
\includegraphics[width=\textwidth]{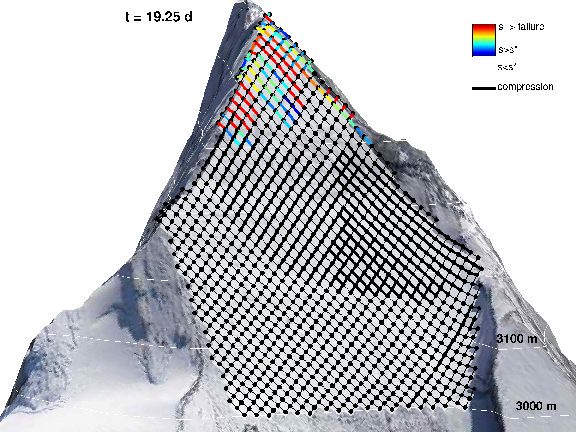}
 \end{minipage}
 \\
 \begin{minipage}{0.48\textwidth}
 \includegraphics[width=\textwidth]{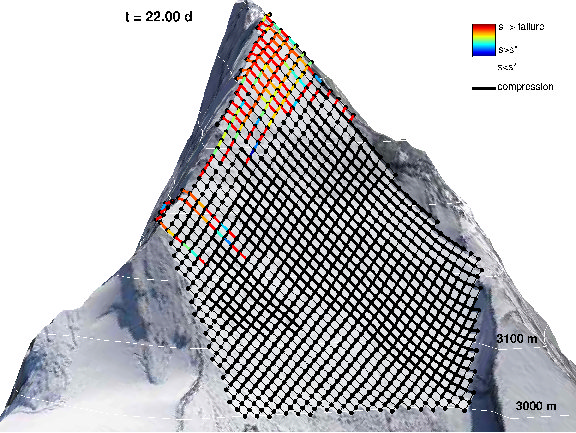}
 \end{minipage}
 &
 \begin{minipage}{0.48\textwidth}
 \includegraphics[width=\textwidth]{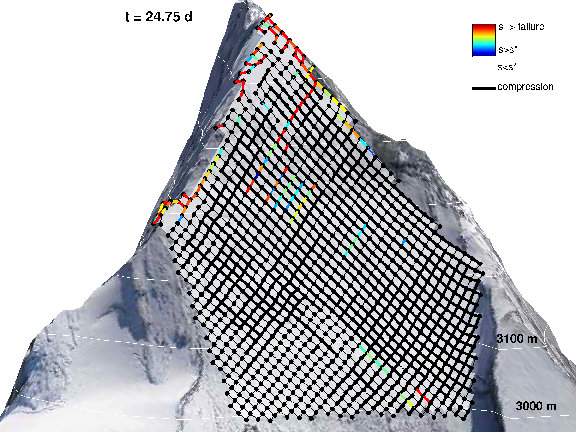}
 \end{minipage}
 \\
 \begin{minipage}{0.48\textwidth}
\includegraphics[width=\textwidth]{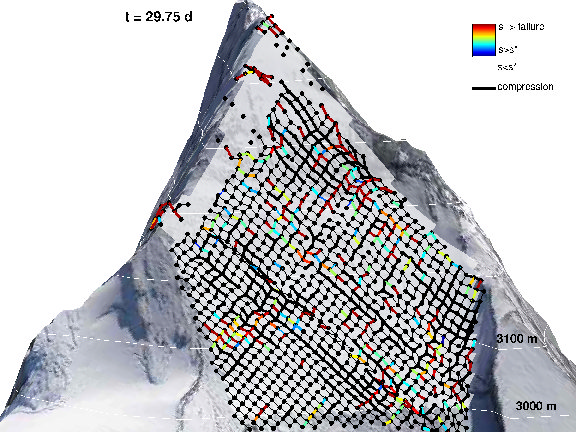}
 \end{minipage}
 &
 \begin{minipage}{0.48\textwidth}
 \includegraphics[width=\textwidth]{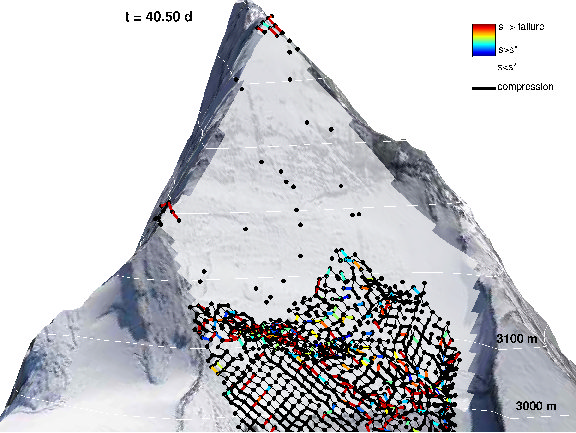}
 \end{minipage} 
 \end{tabular}
 \caption{\label{proginst_masseev}
 Six snapshots describing the rupture 
 progression and sliding instability in the block lattice with
 constant friction coefficient $\mu_0$ for all blocks in the presence of a progressive increase
 the block weights modeling snow/ice addition.  The blocks
 are presented as points at the nodes of the square lattice.
 The color of each bond encodes the time remaining to rupture: red (close to rupture) to blue (far
 from rupture). Bonds in compression are drawn as thick black lines. Bonds without unstable tertiary creep damage are represented as thin black lines. Similar results are obtained
 by progressively decreasing the friction coefficient for all blocks.}
 \end{figure*}

To answer this question, we tested the case of a constant friction
coefficient in the presence of a progressive increase of the block weight.
This means that mass is added at a constant rate on each block.
In all the simulations performed, results show that an
instability starts from the upper part of the glacier, in contradiction
with observation (see Fig. \ref{proginst_masseev}). This could be explained by the bedrock topography: The slope is
indeed steeper in the upper part, inducing an initial
sliding of the upper blocks. Then, this instability propagates downwards and the whole
glacier collapses.

The progressive thickening is therefore not the cause of the 1895 Altels break-off event.

\subsubsection{Is the 1895 Altels break-off due to uniform warming conditions at the
  bedrock?}
At such altitude, the glacier is expected to be cold, i.e. stuck on its bedrock. But we saw that it experienced
successive extreme hot summers, that could have initiated the phenomenon.
 Warming conditions could lead to a lubrication at the bedrock due to melted water penetration and consequent increase of basal water pressure. This can cause decoupling of the glacier with
its bedrock and thus to decrease the friction between ice and bedrock.
There are two ways to model uniform warming conditions in our model,
first by decreasing uniformly the friction coefficient under each block, second by decreasing the tangential component
of the weight of each block.
We tested both approaches which gave very similar qualitative results. 
Again, as expected, the whole glacier
collapses, starting from its upper part (Fig. \ref{proginst_masseev}), for the
same topographical reasons as explained above.

It also seems that differential evolution of the basal properties would affect the stability of this glacier.

\subsubsection{Is the 1895 Altels collapse due to a local increase of water pressure?}
We now show that the {\bf only way} to reproduce the geometry of the rupture is to progressively modify the basal
properties in a restricted area, corresponding to the likely temperate area
at the bedrock \citep{Rothlisberger1981}.

\begin{figure}
\begin{tabular}{c}
\includegraphics[width=0.5\textwidth]{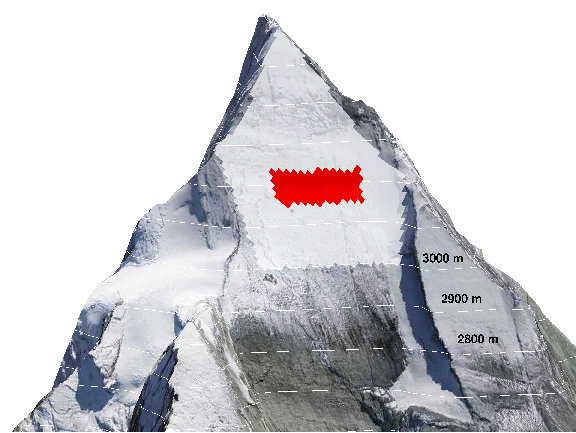}
\\
 \includegraphics[width=0.5\textwidth]{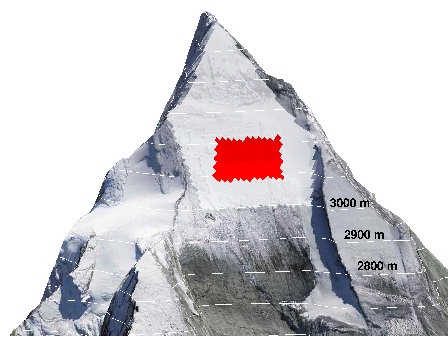}
\\
 \includegraphics[width=0.5\textwidth]{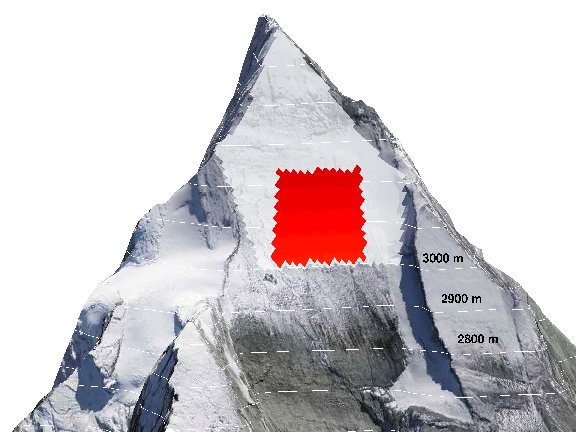} 
\end{tabular}
\caption{\label{zonemuev} Zones where the basal friction coefficient is
  decreased. 
Their geometrical limits were determined according to \citet{Rothlisberger1981}, see Fig. \ref{thermalcondition}.}
\end{figure}

In the following, the friction coefficient is decreased with different
rates $\frac{\delta \mu}{\delta t}$ on three different restricted areas (see Fig. \ref{zonemuev}) corresponding to the assumed temperate
bedrock zone. In this way, we simulate the water penetrating within the glacier leading to a lubrification at the bedrock.

In the following, the value of $\mu_0$ is set larger than $\rm arctan(max(\phi))$, where $\phi$ is the
slope evaluated from the Digital Elevation Model (DEM) of the glacier bed. In this way, the glacier is assumed to be stable when the simulation starts.
At each numerical time step, the friction coefficient
$\mu_0$ is decreased by $\frac{\delta \mu}{\delta t}$.

\subsubsection{Qualitative description of the maturation of rupture in the
  case of a local evolution of basal properties:}
\begin{figure*}[!!!h]
\centering
\begin{tabular}{cc}

\begin{minipage}{0.48\textwidth}
\includegraphics[width=\textwidth]{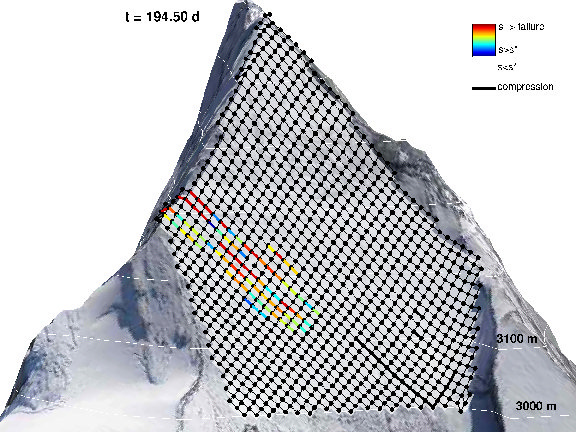}
\end{minipage}
\begin{minipage}{0.48\textwidth}
\includegraphics[width=\textwidth]{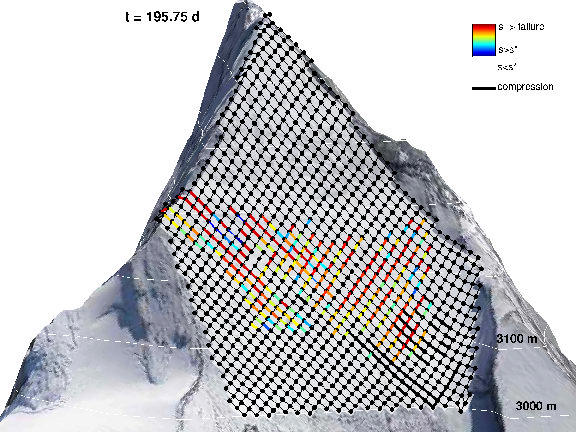}
\end{minipage}
\\
\begin{minipage}{0.48\textwidth}
\includegraphics[width=\textwidth]{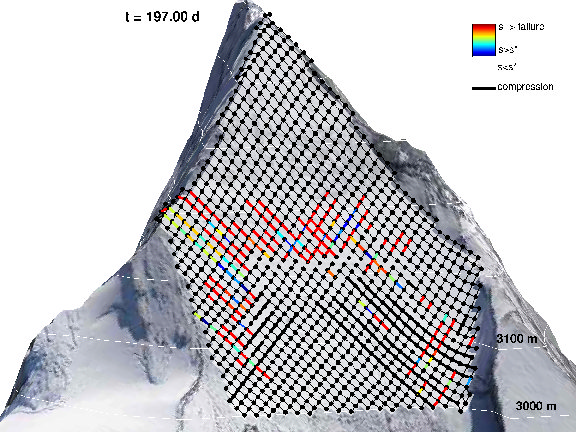}
\end{minipage}
\begin{minipage}{0.48\textwidth}
\includegraphics[width=\textwidth]{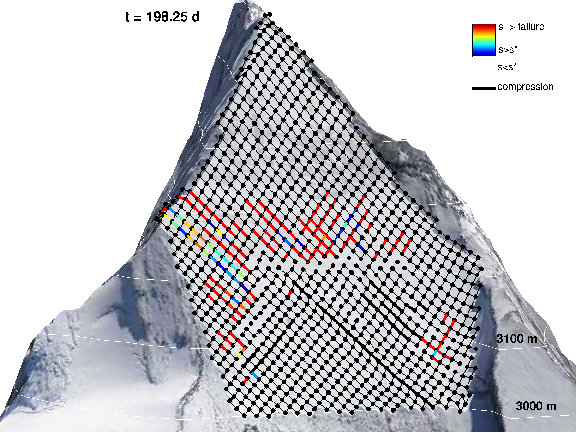}
\end{minipage}
\\
\begin{minipage}{0.48\textwidth}
\includegraphics[width=\textwidth]{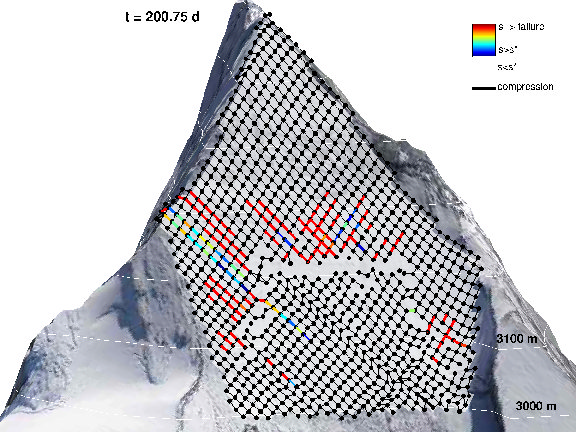}
\end{minipage}
 \begin{minipage}{0.48\textwidth}
\includegraphics[width=\textwidth]{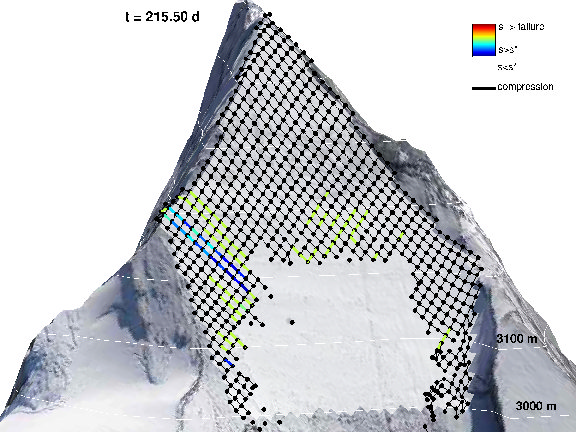}
 \end{minipage}

\end{tabular}
\caption{\label{proginst25}
Six snapshots describing the rupture
progression and sliding instability in the block lattice for the
largest zone where basal friction coefficient was progressively reduced.  The blocks
are presented as points at the nodes of the square lattice.
The color of each bond encodes the time remaining to rupture: red (close to rupture) to blue (far
from rupture). Bonds in compression are drawn as thick black lines. Bonds without unstable tertiary creep damage are represented as thin black lines.}
\end{figure*}

Figure \ref{proginst25} shows the evolution of the set of blocks
in the regime where an instability develops.
Different phases can be distinguished during this simulation:
\begin{enumerate}
\item[(i)] Initially, blocks situated in the area where friction coefficient is
  decreased start sliding where the bedrock slope is largest. Sliding of these blocks lead to
  a change of stress experienced by the bonds. 
This internal bond deformation propagates upstream, as shown in Figure
\ref{proginst25} (1)).
\item[(ii)] Then, the glacier starts accommodating its new stress state,
  resulting in a quiet phase. The number of sliding blocks progressively increases,
leading to stronger interactions and to synchronization of the
sliding blocks  (see Figure \ref{proginst25}, (4) and (5)).
\item[(ii)] After a certain time depending on $\frac{\delta
    \mu}{\delta t}$, the lattice starts fracturing. 
A large crack appears perpendicular to the main
slope around the middle of the lattice  (see Figure \ref{proginst25}, (2) and
(3)). This corresponds to the opening of the crevasse just below the bergschrund (Fig. \ref{Altels1894}). 
\item[(iv)] A global instability develops. The blocks located below the upper
  crevasse start to accelerate and a fracture propagates in the bedrock slope
  direction, forming an unstable slab which finally slides off (see Figure \ref{proginst25}, (6)).  
\end{enumerate}

\clearpage
\subsection{Quantitative results}
\label{quantitativeres}
The results obtained with the spring-block model can be summarized as follows.

\subsubsection{Initiation of the instability}

\begin{figure}
\centering
\begin{tabular}{c}
\includegraphics[width=0.7\textwidth]{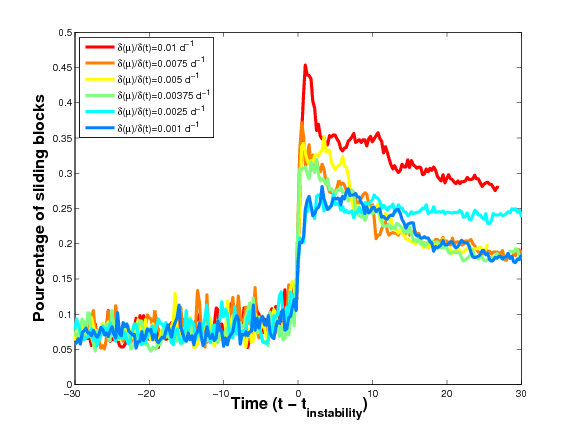}
\\
\includegraphics[width=0.7\textwidth]{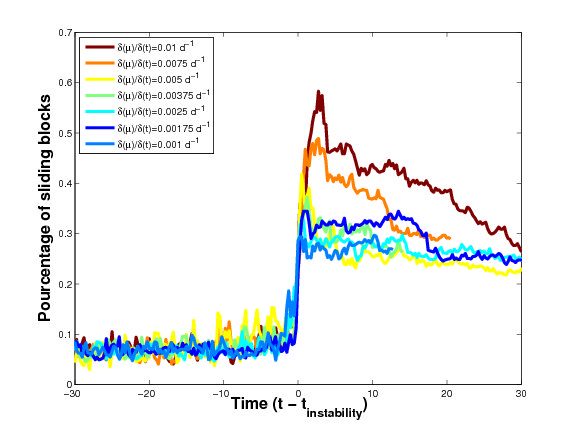}
\\
\includegraphics[width=0.7\textwidth]{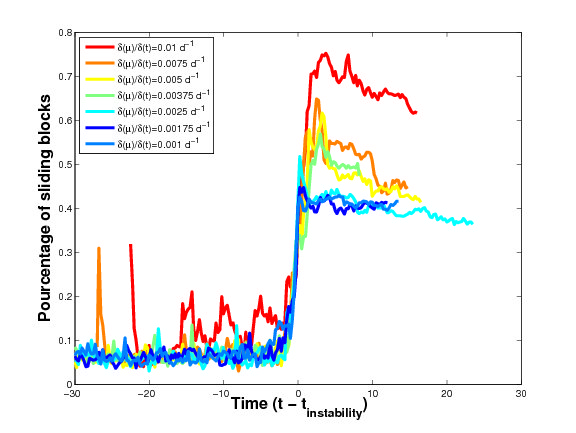}
\end{tabular}
\caption{\label{nslide} Percentage of sliding blocks within the glacier
  for different $\frac{\delta \mu}{\delta t}$ as a function of time, for the
  three different process zones (resp. minimum, medium and maximum
  corresponding to the top to bottom panels of fig. 12).}
\end{figure}

For each of the three process zones (i.e. the area where the friction coefficient is decreased) and for different rates $\delta (\mu) / \delta(t)$ of decrease of the coefficient of friction (RDCF), we have plotted the number of sliding blocks at each time step (see Fig. \ref{nslide}).
It appears that, in all cases, the initiation of the instability occurs very suddenly, typically within one or two days. 
Two different regimes could also be distinguished. A first quiescent one, where isolated blocks slide, and a second active one, where blocks start to slide in clusters leading to the final collapse.

The number of sliding blocks after the onset of the instability depends on the RDFC.
The greater the RDCF, the larger the number of sliding block. 
When the RDCF is small, the glacier has time to adapt to the changes of basal conditions. In this case, the size of the initial unstable cluster is strictly given by the size of the process zone.

\subsubsection{Damage evolution within the glacier}

\begin{figure}
\centering
\begin{tabular}{c}
\includegraphics[width=0.7\textwidth]{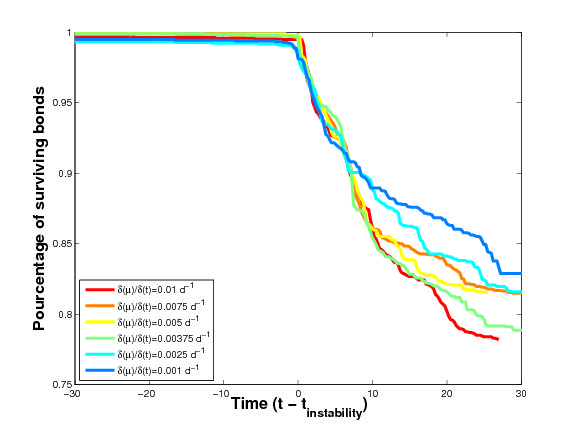}
\\
\includegraphics[width=0.7\textwidth]{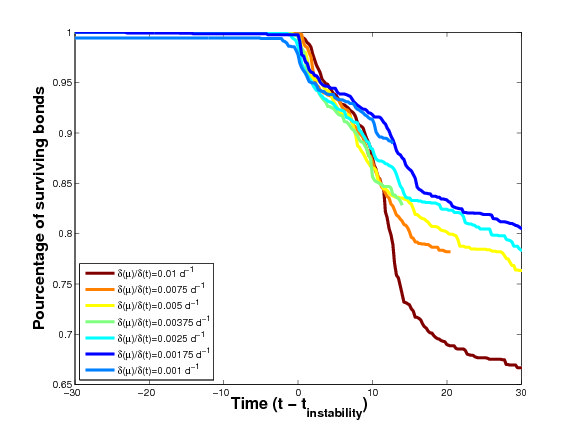}
\\
\includegraphics[width=0.7\textwidth]{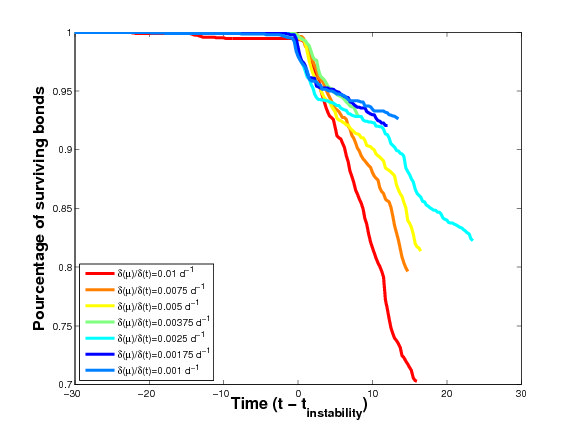}
\end{tabular}
\caption{\label{survbond} Percentage of the surviving bonds within the glacier
  for different rate $\frac{\delta \mu}{\delta t}$ of decrease of the coefficient of friction (RDCF)
  as a function of time, for the
  three different process zones (resp. minimum, medium and maximum corresponding
  to the top to bottom panels of fig. 12).}
\end{figure}

In order to measure the damage evolution within the glacier, we count the
number of surviving bonds during each simulation. 
Fig. \ref{survbond} shows the number of surviving bonds within the lattice at each time step, for different process zones and different rates of decrease of the friction coefficient (RDCF).
A rapid increase of the damage a few days after the initiation of the instability can be observed for all simulations. Moreover, this increase does not seem to depend on the rate of decrease of the friction coefficient. 
Results show that, once the behavior enters in the active regime, bonds start to fail, leading to the openning of the crown crevasse.
This crevasse opens very rapidly, typically in a few days, which was also observed \citep{Heim1895}.

\subsubsection{Energy analysis}

\begin{figure}
\begin{center}
\includegraphics[width=0.7\textwidth]{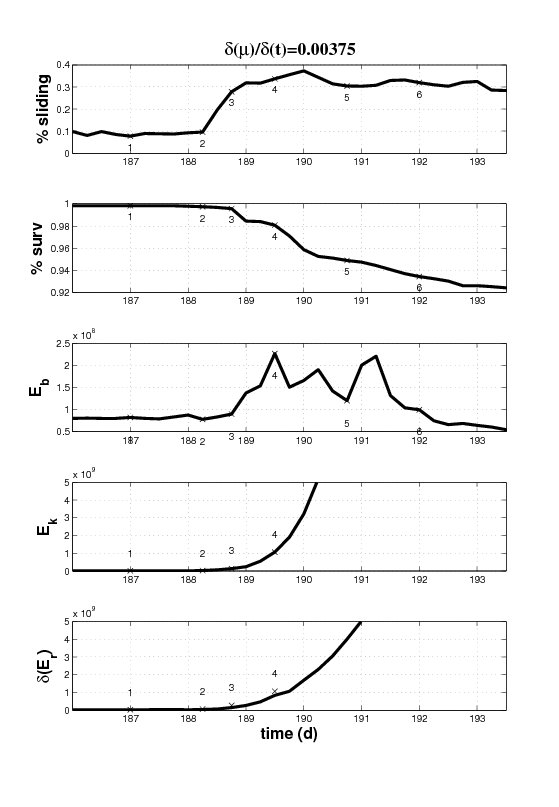}
\end{center}
\caption{\label{compE} \sf
Evolution of the energy stored in the bonds, of the kinetic energy and of the
radiated energy during the destabilization process.}
\end{figure}
In order to detect precursors to the rupture, we plotted the evolution of the energy stored in the bonds, the kinetic energy and the
radiated energy during the destabilization
process.

A typical result is shown in Fig. \ref{compE}. Six phases can be distinguished.
(1) The glacier is still in a stable phase. (2) Initiation of the instability. (3) The number of sliding blocks drastically increases. (4) The number of sliding blocks reaches a maximum and the energy stored in the bonds increases. Note that the increase of kinetic and radiated energies starts to be visible. (5) The energy stored in the bonds drops because the crown crevasse opens up. The number of sliding blocks remains unchanged and the instability is now established. Both the kinetic and radiated energies increase. (6) After an increase of the energy stored in the bonds, cluster of bonds are failing.

These results show that no precursor of the instability can be inferred from the time evolution of the 
different types of energy more than a few days before the break-off.

\subsubsection{Occurrence of the instability as a function of $\delta(\mu)/\delta(t)$. }
\begin{figure}
\begin{center}
\includegraphics[width=0.7\textwidth]{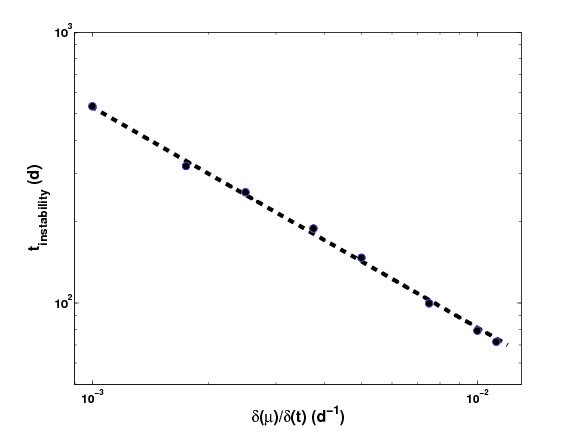}
\end{center}
\caption{\label{mutfex} Time of rupture as a function of
the RDCF  $\delta(\mu)/\delta(t)$, for the medium process zone.
The dotted line plots the equation: $t_i \sim (\frac{\delta(\mu)}{\delta(t)})^{-0.82}$
}
\end{figure}

To assess if the rate of decrease of the friction coefficient (RDFC: $\delta(\mu)/\delta(t)$) influences the final time of rupture, we performed independent runs with different rates and evaluated their respective rupture times.
The results show that the rupture occurs earlier for greater RDFC, which is not surprising (see Fig. \ref{mutfex}).
However, the time of rupture does not depend linearly on RDFC but follows a power-law with an exponent of -0.82.
This means that, for small RDFC, the glacier has time to adapt to these changes and the final instability arises later than in the case of large RDFC.

\begin{figure}
\begin{center}
\includegraphics[width=0.7\textwidth]{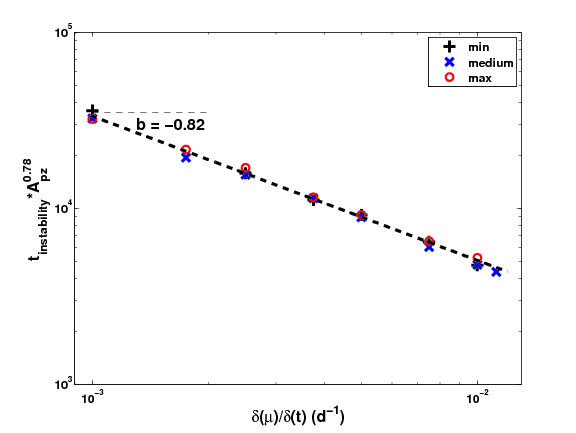}
\end{center}
\caption{\label{mutf} Time of rupture as a function of
  $\delta(\mu)/\delta(t)$.
The dotted line plots the equation: $t_i \cdot A^{0.78} \sim 116 \cdot (\frac{\delta(\mu)}{\delta(t)})^{-0.82}$
}
\end{figure}

The influence of the area of the process zone was tested and 
we found an inverse effect compared with the RDFC. Specifically,
a glacier for which a large RDFC acts on a small process zone becomes unstable 
after the same time as a glacier subjected to a small RDFC applied to a large process zone.
This can be summarized by plotting the reduced variable $t_c*A^{0.78}$ as a function 
of $(\frac{\delta(\mu)}{\delta(t)})^{-0.82}$, as shown in Fig. \ref{mutf}.

Unfortunately, 
in a real situation, the a priori determination of these parameters (RDFC and process zone area) is far from being possible, especially for the area of the process zone.

\section{Conclusion}
The Altels glacier fall of 1895 is the
largest ice fall known in the Alps. The mechanisms leading to this event are not fully understood. With a new model developed by \citet{faillettaz&al2009} of the progressive maturation
of a heterogenous mass towards a gravity-driven instability, characterized 
by the competition between frictional sliding
and tension cracking, we have contributed to a better understanding of this event. We used an array of sliding blocks on an inclined
(and curved) basal topography, which interacts 
via elastic-brittle springs. A realistic state- and rate-dependent friction
law was used for the block-bed interaction.
We modeled the inner material properties of the mass and its progressive damage eventually
leading to failure, by means of a laboratory-based stress corrosion law governing
the rupture of the springs. 

Our simulations showed that the only way to reproduce the particular arch shape of the
crown crack of the Altels glacier fall was to reduce the basal friction coefficient
in a limited area. 
Such a break-off arises because of the onset of a
weak zone at the interface between the glacier and its bedrock, probably due
to melted water infiltrations trapped within the glacier. 
Climatic observations before the collapse
support this assumption as they showed hot summers before the instability, explaining such melted water
infiltrations.

Moreover, a two-step behavior could be evidenced in our simulations: (i) a first quiescent phase, without visible changes with a duration depending on the
rate of decrease of the friction coefficient (RDFC). (ii) An active regime with a rapid increase of basal
motion within a few days before the break-off. As a consequence, a crown crevasse opens a few days later (which was observed) and the final rupture occurs. 
This means that the
destabilization process of a hanging glacier due to a progressive warming of
the ice/bed interface towards a temperate regime is expected to occur without 
visible signs until a few days prior to the collapse.

The area of the process zone and the RDFC have an equivalent relative influence on the time of the onset on the instability. A small area of process zone with a large RDFC would lead to the same behavior as a large area of process zone with a small RDFC.
From a practical point of view, the knowledge of both parameters is needed to predict the onset of such an instability. The determination of these two parameters is not yet possible, especially for the area of the process zone.
The collapse of three power laws shown in Fig. \ref{mutf} can be rewritten 
as $t_i \sim (A \cdot (\delta \mu / \delta t))^{-\nu}$, where $\nu \approx 0.8$.
This law expresses a combined ``size'' effect (through the term $A$)
and a rate-dependence effect (through the term $\delta \mu / \delta t$), which 
is also found qualitatively similar in most heterogenous mechanical systems going to failure
\citep{Carpinteri,Collins}.
Interestingly, the combined dependence of the failure time $t_i$
on the unstable area $A$ and on the rate $\delta \mu / \delta t$ is
through their product, suggesting that the driving mechanism 
for the failure time is the total shear force applied to the unstable area.

A recent paper \citet{faillettaz&al2010} showed that seismic measurements could help predict 
the approaching mechanical instabilities of cold hanging glaciers with the help of some seismic precursors 
(e.g. by using changes in the statistical
behavior of icequakes), before the instability becomes obvious by its visual impacts. Unfortunately, we could not find any seismic precursors for instabilities driven by infiltrated melted water.

In a more general context, global climate warming may influence the
stability of cold hanging glaciers. Moreover, as the rupture process takes
some time to develop and external precursors are only visible a few days prior to the break-off, some cold hanging glaciers
could already be in the unstable phase where the instability is
already developing. 
An early warning of such events is still far from beeing possible.

\end{document}